\documentclass[
preprint,
]{revtex4-1}

\usepackage{graphicx}
\usepackage{dcolumn}
\usepackage{bm}

\begin{document}


\title{Neutrinoless Double Beta Decay and Light Sterile Neutrino}

\author{C. H. Jang}
\author{B. J. Kim}
\author{Y. J. Ko}
\author{K. Siyeon}\email{siyeon@cau.ac.kr}
\affiliation{Physics Department, Chung-Ang University, Seoul 06974 Korea}

\date{November 25, 2018}

\begin{abstract}
Recent neutrino experiment results show a preference for the normal neutrino mass ordering.
The global efforts to search for neutrinoless double beta decays undergo a broad gap with the approach to the prediction in the three-neutrino framework based on the normal ordering.
This research is intended to show that it is possible to find a neutrinoless double beta decay signal even with normal ordered neutrino masses.
We propose the existence of a light sterile neutrino as a solution to the higher effective mass of the electron neutrino expected by the current experiments.
A few short-baseline oscillation experiments gave rise to a limit on the mass of the sterile neutrino and its mixing with the lightest neutrino.
We demonstrate that the results of neutrinoless double beta decays can also narrow down the range of the mass and the mixing angle of the light sterile neutrino.

\begin{description}
\item[PACS numbers]
13.15.+g, 14.60.Pq, 14.60.St
\item[Keywords]
neutrinoless double-beta decay, Majorana phase, sterile neutrino, AMoRE
\end{description}
\end{abstract}

\maketitle

\section{\label{sec:level1}Introduction}

Search for neutrinoless double-beta decay($0\nu$DBD) is the experiment that is running the most in the recent years, and it is currently the only way to tell if neutrinos are Dirac or Majorana particles \cite{Giunti_Kim:2009}\cite{Vergados:2002pv}. To clearly differentiate it from the double-beta decay(DBD or $2\nu$DBD), which naturally emits two neutrinos, we will express that rare decay as $0\nu$DBD\cite{Barea:2012zz}\cite{Bilenky:2012qi}. The seesaw model which has been developed to explain the light neutrino masses and the leptogenesis model which is considered as a solution to the matter-antimatter asymmetry were built on the idea that neutrinos are Majorana particles\cite{lep}-\cite{Luty:un}. In order to justify the high-energy models of the seesaw mechanism and the leptogenesis, a low-energy approach such as the neutrinoless double beta decay is essential \cite{Endoh:2002wm}\cite{Siyeon:2016wro}. The following experiments are examples of $0\nu$DBD with the limit on the effective electron-neutrino mass on the order of 0.1 eV - 1 eV : KamLAND-Zen\cite{KamLAND-Zen:2016pfg}, EXO-200\cite{Albert:2017owj} CUORE\cite{Alduino:2017ehq}, GERDA\cite{Agostini:2018tnm}, Majorana\cite{Alvis:2018yte} and NEMO3\cite{Arnold:2018tmo}. Advanced Mo-based Rare process Experiment(AMoRE)\cite{Alenkov:2015dic}\cite{Lee:2016pmi}\cite{Luqman:2016okt} is a search for $0\nu$DBD with the highest sensitivity of Mo-100 isotope. Depending on the background level, target mass and running period, AMoRE consists three stages: pilot, phase 1, and phase 2, aiming for the sensitivity of effective electron neutrino mass of 0.21-0.40 eV, 0.07-0.14 eV, and 0.012-0.022 eV, respectively\cite{Alenkov:2015dic}. In this work, those AMoRE stages are used for stepwise comparison with Majorana mass models.

The $0\nu$DBD strongly depends on the type of mass ordering because the two vertices of the double-beta decays are connected by a virtual Majorana electron neutrino. The effective mass $\langle m_{\beta\beta}\rangle$ of the virtual electron neutrino, which has a significant contribution from $m_1$, varies depending on whether the masses follow the normal ordering(NO) or the inverted ordering(IO). Recent results from long-baseline neutrino experiments, NOvA\cite{Adamson:2017gxd} and T2K\cite{Abe:2018wpn}, support the NO and non-zero CP violation. In case of NO hierarchy, $m_1 \ll m_2 \ll m_3$, the measurement of $\langle m_{\beta\beta}\rangle$ requires a resolution of at least 0.004 eV. Current efforts to measure $\langle m_{\beta\beta}\rangle$ described in Refs. \cite{KamLAND-Zen:2016pfg}-\cite{Alenkov:2015dic} and even their future projects have difficulties to approach such a low limit. Here, we propose the existence of a fourth sterile neutrino as a solution to the conflict between the sensitivities in those experiments and neutrino masses of the NO hierarchy. We also consider the contribution of $m_4$ to $\langle m_{\beta\beta} \rangle$, and track backwards the implication of the measured $\langle m_{\beta\beta} \rangle$ values to the bound on $m_4$.  The search for a sterile neutrino has been performed by short-baseline oscillation experiments\cite{Adamson:2016jku}-\cite{Almazan:2018wln}. They narrow down the allowed values of $\Delta m_{14}^2$ vs. $\sin^22\theta_{14}$ by excluding the regions scanned by the experiments. We show that the results of neutrinoless double beta decays can also provide exclusion curves in the plane of $\Delta m_{14}^2$ vs. $\sin^22\theta_{14}$.

The outline is as follows: Section II describes the relation between the effective mass determined by the half-life of $0\nu$DBD and the neutrino masses. Section III extends the sensitivity of the effective mass with a fourth neutrino and discuss its implication to the sterile neutrino searches. Section IV concludes the work by summarizing and discussing the results.

\section{Neutrinoless double beta decay}

The half life of a double-beta decay, $\mathcal{N}(\mathrm{A,Z}) \rightarrow \mathcal{N}(\mathrm{A,Z+2}) + 2e^- + 2\overline{\nu}_e$, is given by
\begin{eqnarray}
(T^{2\nu}_{1/2})^{-1}=G_{2\nu}|\mathcal{M}_{2\nu}|^2,
\end{eqnarray}
with a phase factor $G_{2\nu}$ and the nuclear matrix element $\mathcal{M}_{2\nu}$ \cite{Elliott:1987kp}-\cite{Saakyan:2013yna}. If neutrinos are Majorana fermions, lepton number violating $(\Delta L=2)$ processes are allowed in such a way that only two electrons are emitted without antineutrinos. The $0\nu$DBD, $\mathcal{N}(\mathrm{A,Z}) \rightarrow \mathcal{N}(\mathrm{A,Z+2}) + 2e^-$, has a decay width suppressed by the squared effective mass $\langle m_{\beta\beta}\rangle^2$,
\begin{eqnarray}
(T^{0\nu}_{1/2})^{-1}=G_{0\nu}|\mathcal{M}_{0\nu}|^2\langle m_{\beta\beta}\rangle^2,
\end{eqnarray}
where $G_{0\nu}$ and $\mathcal{M}_{0\nu}$ are the phase space factor and the nuclear matrix element of the neutrinoless decay, respectively \cite{Bilenky:2012qi}\cite{Faessler:2012ku}.
Two Majorana neutrinos produced in double-beta decay are connected to each other producing an internal line between the two vertices. The virtual neutrino between the two weak vertices has an effective mass defined by
\begin{eqnarray}
\langle m_{\beta\beta}\rangle=|\sum_{i=1}^3 U_{ei}^2m_i|. \label{effective}
\end{eqnarray}

\begin{figure}
\resizebox{75mm}{!}{\includegraphics[width=0.75\textwidth]{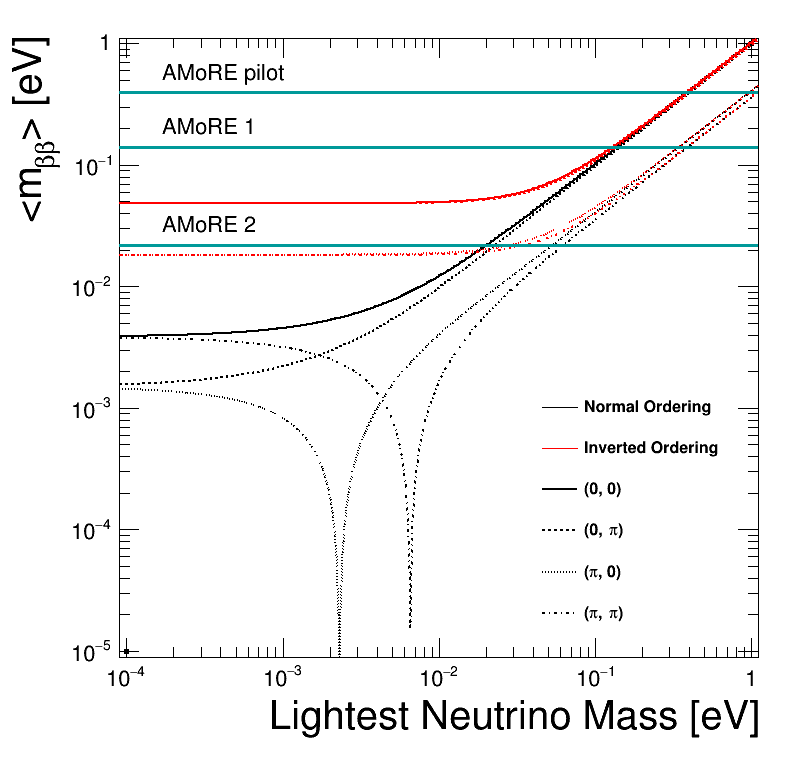}}
\caption{\label{fig_Mbb_light}$\langle m_{\beta\beta} \rangle$ vs. $m_{\mathrm{light}}$ for three neutrinos. Two Majorana phases $\alpha_1$ and $\alpha_2$ independently run from 0 to $2\pi$. Each boundary curve is obtained by the specified assignment of $(\alpha_1,\alpha_2)$.  }
\end{figure}

The electron neutrino which couples to the electron is expressed as a superposition of the three massive neutrinos, such that
\begin{eqnarray}
 \nu_e = \sum_{i=1}^3 U_{ei} \nu_i.
\end{eqnarray}
The unitary transformation matrix of Majorana neutrinos $U$ includes two Majorana phases in addition to the three mixing angles and the Dirac phase of the PMNS matrix, such that $U= U_{PMNS} Diag(1, e^{\varphi_1/2}, e^{\varphi_2/2})$. The effective mass $\langle m_{\beta\beta}\rangle$ can be estimated using the PMNS matrix and the mass-squared differences $\Delta m^2_{31}$ and $\Delta m^2_{21}$, while the Majorana phases $\varphi_1$, $\varphi_2$ and $m_1$ are unconstrained.  Fig.\ref{fig_Mbb_light} shows the dependency of $\langle m_{\beta\beta}\rangle$ on the lightest neutrino mass, when the full range between 0 and 2$\pi$ is allowed for $\varphi_1$ and $\varphi_2$. The lightest mass in normal ordering(NO) is $m_1$ while the one in inverted ordering(IO) is $m_3$.
The sensitivities of operating or planed double-beta decay experiments are far worse than the sensitivity required to measure $0\nu$DBD if NO is correct, as shown in Fig.\ref{fig_Mbb_light}. Thus, we consider the existence of a sterile neutrino as an explanation to possible neutrinoless double-beta decay events in those experiments.

All figures hereafter are obtained assuming the following values for the neutrino mixing parameters\cite{Tanabashi:2018oca}:
\begin{eqnarray}
  && \begin{array}{lclc}
        \Delta m_{21}^2 &=& 7.54 \times 10^{-5}~\rm{(eV)}^2 & \\
        \sin^2\theta_{12} &=& 3.08 \times 10^{-1} &
  \end{array} ~\rm{NO ~or ~IO} \label{best_fit} \\ \nonumber \\
  && \begin{array}{lclc}
        \Delta m_{31}^2 &=& 2.43\times10^{-3}~\rm{(eV)}^2 & \\
        \sin^2\theta_{13} &=& 2.34\times10^{-2} & \\
        \sin^2\theta_{23} &=& 4.37\times10^{-1} &
  \end{array} ~\rm{NO} \label{best_fit_no} \\ \nonumber \\
  && \begin{array}{lclc}
        \Delta m_{31}^2 &=& 2.38 \times 10^{-3}~\rm{(eV)}^2 &\\
        \sin^2\theta_{13} &=& 2.40 \times 10^{-2} &\\
        \sin^2\theta_{23} &=& 4.55 \times 10^{-1} &
      \end{array}  ~\rm{IO} \label{best_fit_io}
\end{eqnarray}
Regarding the Dirac CP phase $\delta_1$, recent observations of long-baseline oscillations by NOvA\cite{Adamson:2017gxd} and T2K\cite{Abe:2018wpn} obtained a preferred value of $\delta_1=-\pi/2$. Following the convention from Ref.\cite{Giunti:2015kza}, the effective electron-neutrino mass in Eq.(\ref{effective}) is
\begin{eqnarray}
\langle m_{\beta\beta}\rangle=|\sum_{i=1}^3 \mu_i e^{-\imath\alpha_{i-1}}|, \label{effective3mu}
\end{eqnarray}
where $\mu_i \equiv m_i|U_{ei}|^2$ and $\alpha_1=\varphi_1$, and $\alpha_2=2\delta_1+\varphi_2$. As shown in Fig.\ref{fig_Mbb_light}, the contribution of the Dirac phase does not affect the results when Majorana phases $\alpha_i$ run from 0 to $2\pi$. The plots $\langle m_{\beta\beta} \rangle$ vs. $m_{\mathrm{light}}$ are drawn using the fixed values in Eqs.(\ref{best_fit})-(\ref{best_fit_io}). The red curves are obtained by the values for IO in Eq.(\ref{best_fit}) and Eq.(\ref{best_fit_io}), while the grey curves are obtained by the values for NO in Eq.(\ref{best_fit}) and Eq.(\ref{best_fit_no}). When the lightest mass is lower than $10^{-2}$ eV, the mass ordering can be considered to be hierarchical. The expressions like normal hierarchy(NH) and inverted hierarchy(IH) are also used to specify a type of mass. Determining the absolute neutrino mass scale is challenging, as it requires a precise measurement of the tritium decay spectrum.\cite{Drexlin:2005zt}.

For IO, the upper bound in Fig.\ref{fig_Mbb_light} is obtained by $|\mu_1+\mu_2\pm\mu_3|$ with $(\alpha_1, \alpha_2)=(0,0)$ and $(0,\pi)$, respectively, while the lower bound is obtained by $|\mu_1-\mu_2\pm\mu_3|$ with $(\alpha_1, \alpha_2)=(\pi,0)$ and $(\pi,\pi)$.
For NO, the bounds in Fig.\ref{fig_Mbb_light} are determined by the following four combinations of $\mu_i$'s, such that
\begin{eqnarray}
&& |\mu_1\pm\mu_2\pm\mu_3|, \label{++} \\
&& |\mu_1\mp\mu_2\pm\mu_3|, \label{+-}
\end{eqnarray}
which indicate the four values of $(\alpha_1, \alpha_2)$ for NO shown in Fig.\ref{fig_Mbb_light}. The two expressions in Eq.(\ref{++}) approach to each other, as $m_1$ goes to zero. The same applies the two expressions in Eq.(\ref{+-}). The five-year AMoRE target sensitivity to $\langle m_{\beta\beta} \rangle$ is 0.07-0.14 eV, and 0.012-0.022 eV for the phase 1 and phase 2, respectively. AMoRE pilot may reach a sensitivity 0.21-0.40 eV with a three-year measurement\cite{Alenkov:2015dic}. Only the upper bounds on the sensitivities are shown in the figures.
\begin{figure}
\resizebox{75mm}{!}{\includegraphics[width=0.75\textwidth]{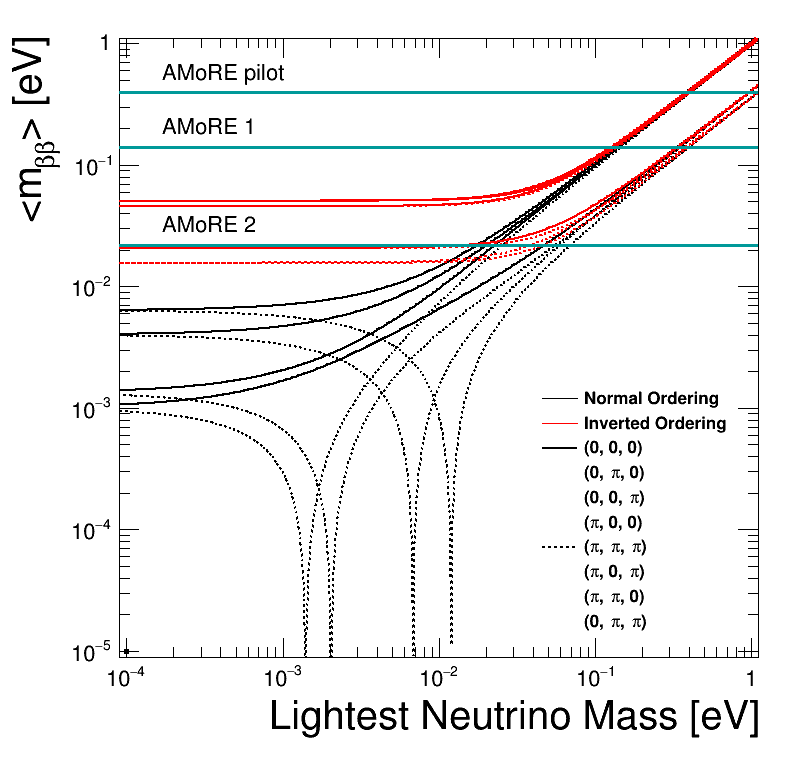}}
\caption{\label{fig_Mbb_light4}$\langle m_{\beta\beta} \rangle$ vs. $m_{\mathrm{light}}$ for 3+1 neutrinos. All three phases $\alpha_1, \alpha_2$ and $\alpha_3$ independently run from 0 to $2\pi$, assuming $\sin^22\theta_{14}=0.01$ and $\Delta m^2_{41}=1\mathrm{eV}^2.$ As the lightest mass approaches zero, the solid and the dashed curves pairwise approach each other. From top to bottom, the solid curves represent the choices $(\alpha_1, \alpha_2, \alpha_3)=(0,0,0),(0,\pi,0),(0,0,\pi)$, and $(\pi,0,0)$, while the dashed curves represent the choices $(\alpha_1, \alpha_2, \alpha_3)=(\pi,\pi,\pi),(\pi,0,\pi),(\pi,\pi,0)$, and $(0,\pi,\pi)$.}
\end{figure}

\section{$\langle m_{\beta\beta} \rangle$ with the 4th neutrino}

The 3+1 neutrino model consists of the three Standard Model(SM) neutrinos $(\nu_e, \nu_\mu, \nu_\tau)$ and one sterile neutrino $\nu_s$. The virtual neutrino $\nu_e$ connecting the two vertices in the double-beta decay is a linear combination of the four massive neutrinos $(\nu_1, \nu_2, \nu_3, \nu_4)$.
\begin{eqnarray}
 \nu_e = \sum_{k=1}^4 U_{ek} \nu_k.
\end{eqnarray}
The $4\times 4$ mixing matrix $U$ of Majorana neutrinos is given by $U= \widetilde{U} Diag(1, e^{\varphi_1/2}, e^{\varphi_2/2}, e^{\varphi_3/2})$ with
\begin{eqnarray}
    \widetilde{U} &=& R_{34}(\theta_{34},\delta_3)R_{24}(\theta_{24},\delta_2)R_{14}(\theta_{14}) \cdot \nonumber \\
    & & \cdot R_{23}(\theta_{23})R_{13}(\theta_{13},\delta_1)R_{12}(\theta_{12}),
    \label{4by4trans}
\end{eqnarray}
where $R_{ij}$ indicates a single angle rotation in $i-j$ plane. The elements of the first row of $\widetilde{U}$ are $ \{ \widetilde{U}_{ek} \} = \{c_{14}c_{13}c_{12},~c_{14}c_{13}s_{12},~c_{14}s_{13}e^{-\imath\delta_1},~s_{14} \}.$

\begin{figure*}
\resizebox{170mm}{!}{\includegraphics[width=.55\textwidth]{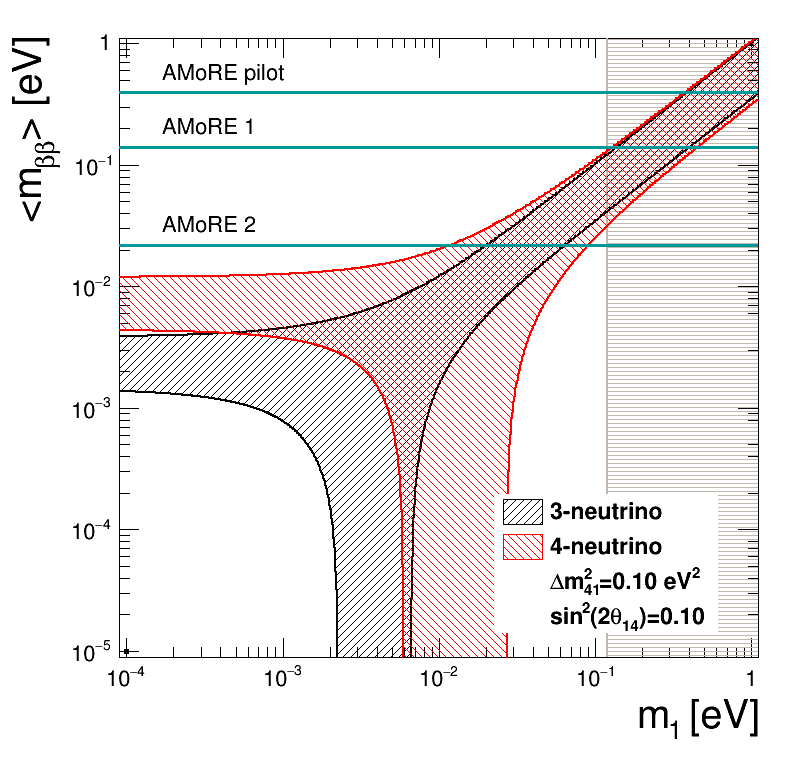}
                    \includegraphics[width=.55\textwidth]{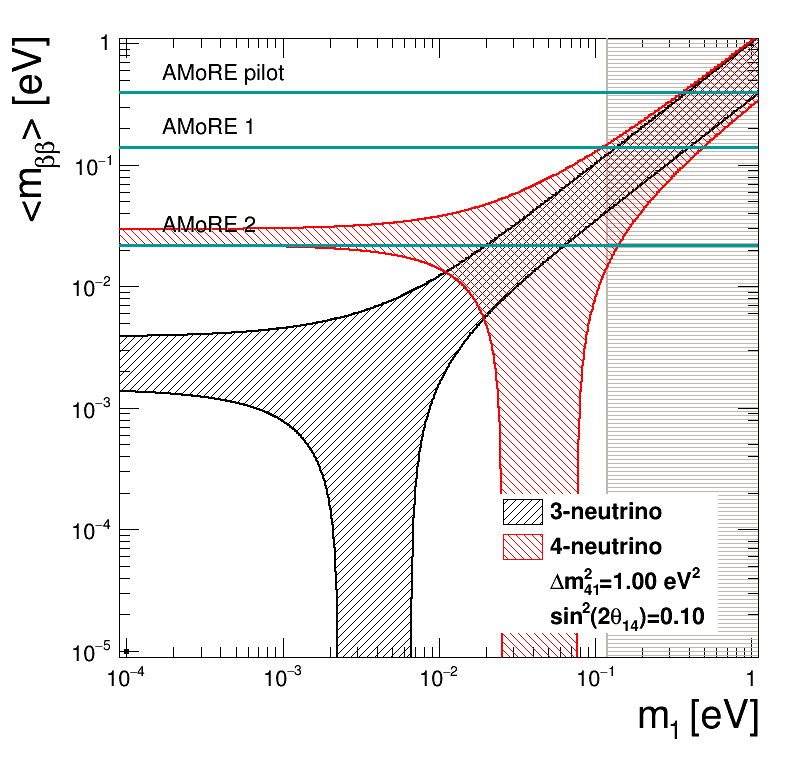}
                    \includegraphics[width=.55\textwidth]{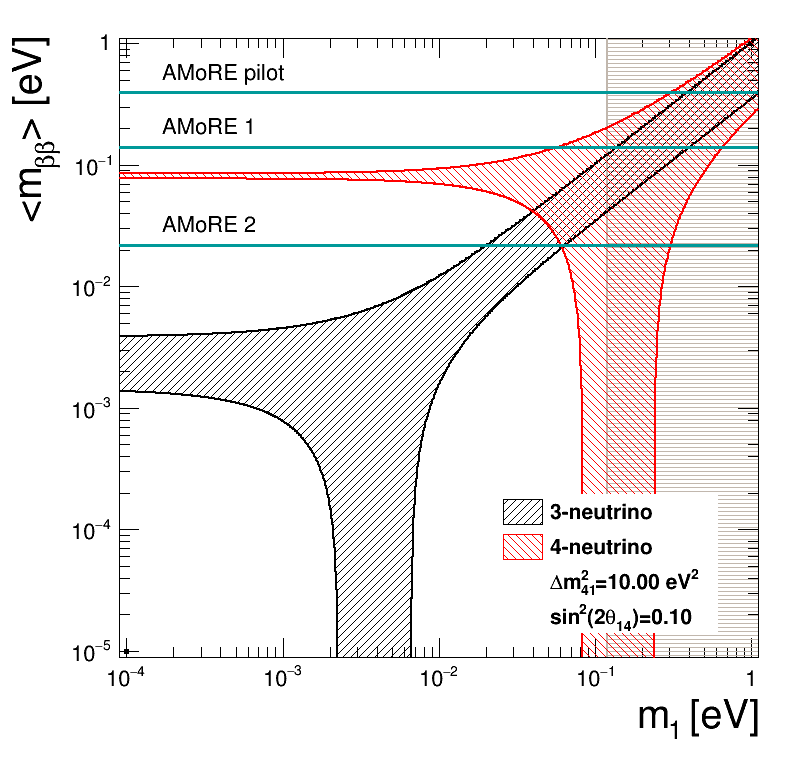}}
\caption{\label{fig_nh_4neutrino_bj1}$\langle m_{\beta\beta} \rangle$ vs. the lightest neutrino mass $m_1$ for NO. Mixing angle is fixed at $\sin^22\theta_{14}=0.1$. Three different values of $\Delta m_{41}^2$, $0.1~\rm{eV}^2, ~1~\rm{eV}^2$ and $10~\rm{eV}^2$ are shown in the three panels from left to right. The values of $m_1$ above $0.12~\rm{eV}$ are excluded by cosmological observations. }
\end{figure*}

The effective electron-neutrino mass in Eq.(\ref{effective}) is now extended to the one with four massive neutrinos as follows;
\begin{eqnarray}
    \langle m_{\beta\beta}\rangle &=& |\sum_{k=1}^4 m_k |U_{ek}|^2 e^{-\imath\alpha_{k-1}}|,
    \nonumber \\
    &\equiv &
    |\mu_1+\mu_2e^{-\imath\alpha_1}+\mu_3e^{-\imath\alpha_2}+\mu_4e^{-\imath\alpha_3}| \label{effective4m}
\end{eqnarray}
where $\alpha_1=\varphi_1, ~\alpha_2=2\delta_1+\varphi_2$, and $\alpha_3=\varphi_3$. The $\alpha_i$'s can take values from 0 to $2\pi$, independent of the Dirac phase. The curves in Fig.\ref{fig_Mbb_light4} are obtained by fixing $\alpha_i$ to be 0 or $\pi$. For any values of $\alpha_i$'s, the line in the plot would fall between the two extremes, as shown in Fig.\ref{fig_nh_4neutrino_bj1}.
\begin{figure*}
\resizebox{170mm}{!}{\includegraphics[width=.55\textwidth]{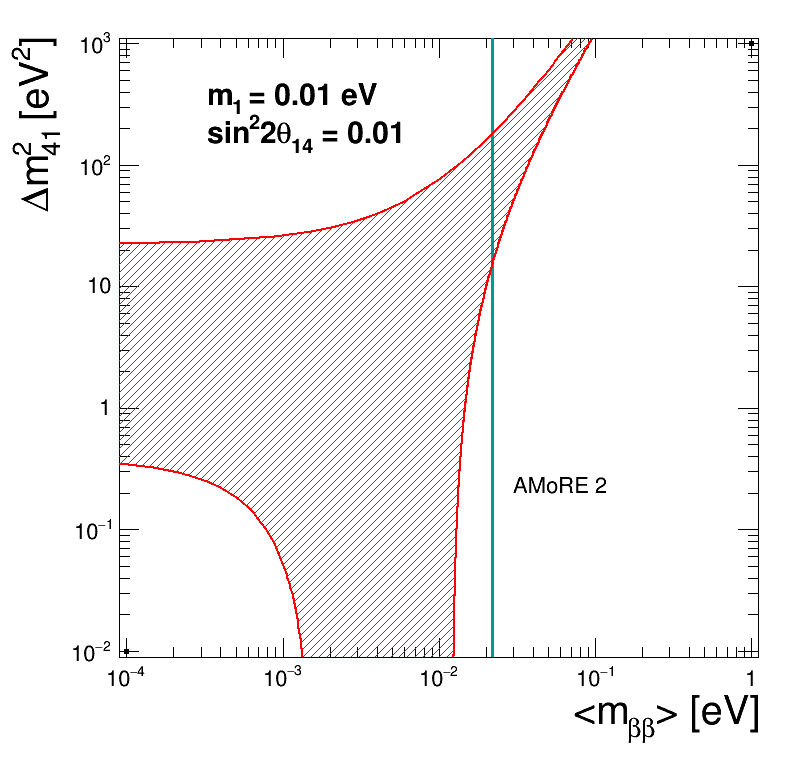}
                    \includegraphics[width=.55\textwidth]{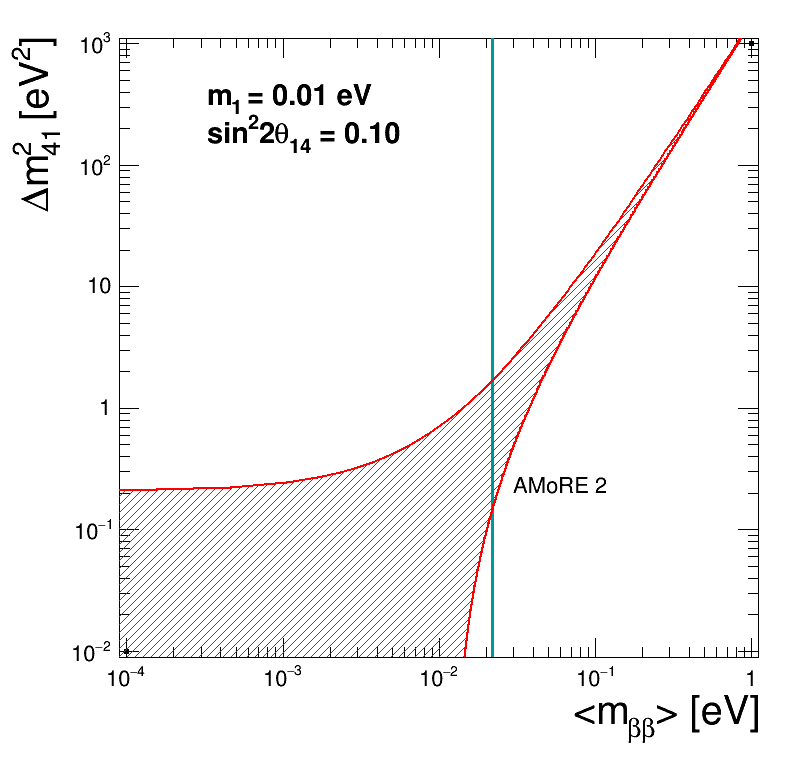}
                    \includegraphics[width=.55\textwidth]{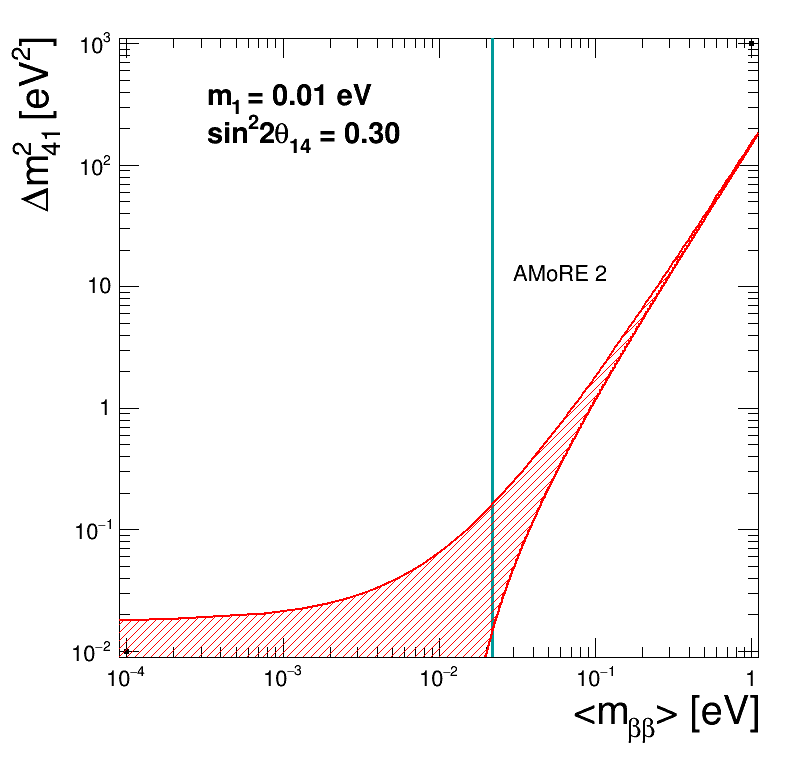}}
\caption{\label{fig_nh_4neutrino_bj2}$\Delta m^2_{41}$ vs. the effective mass $\langle m_{\beta\beta} \rangle$ for NH.
The lightest neutrino mass is fixed at $m_1=0.01~$eV. Three different values of $\sin^22\theta_{14}$, 0.01, 0.10, and 0.30 are shown in the three panels from left to right. From the measurement of $\langle m_{\beta\beta} \rangle$, one can estimate the allowed range of $\Delta m^2_{41}$ which enhances the neutrinoless double-beta decay rate.}
\end{figure*}

Whereas the $\langle m_{\beta\beta}\rangle$ is understood in terms of $\alpha_1, \alpha_2$ and $m_\mathrm{light}$ in the three-neutrino theory, the interpretation of $\langle m_{\beta\beta} \rangle$ in the four-neutrino theory is rather complex due to the additional free parameters, $m_4, \theta_{14},$ and $\alpha_3$. Fig.\ref{fig_Mbb_light4} shows typical plots with three Majorana phases running from 0 to $2\pi$ assuming fixed values for $(\Delta m_{41}^2, \sin^2 2\theta)$. Recent results from short-baseline neutrino oscillation experiments help narrow down the allowed range of $m_4$, as well as the range of $\theta_{14}$. Then, the $\langle m_{\beta\beta} \rangle$ curves in Fig.\ref{fig_Mbb_light4} consist of the following four pairs
\begin{eqnarray}
&& |\mu_1\pm\mu_2\pm\mu_3\pm\mu_4| \label{+++} \\
&& |\mu_1\pm\mu_2\mp\mu_3\pm\mu_4| \label{+-+} \\
&& |\mu_1\pm\mu_2\pm\mu_3\mp\mu_4| \label{++-} \\
&& |\mu_1\mp\mu_2\pm\mu_3\pm\mu_4| \label{-++}
\end{eqnarray}
from top to bottom. In each pair, the upper signs correspond to the solid curves, while lower signs correspond to the dashed curves in Fig.\ref{fig_Mbb_light4}.

Fig. \ref{fig_nh_4neutrino_bj1} shows how the existence of the 4th neutrino changes the accessible range of effective mass. Limiting the neutrino masses in normal hierarchy, three values of mass-squared difference $\Delta m_{41}^2$, $0.1~\rm{eV}^2$, $1~\rm{eV}^2$ and $10~\rm{eV}^2$, are shown for $\sin^22\theta_{14}=0.1$. When the lightest mass is large such that $m_1\sim m_2\sim m_3$ which is called degenerate or quasi-degenerate, the plots are shown with $\langle m_{\beta\beta} \rangle \sim m_1$. Furthermore, the value of the lightest mass above $0.12~eV$ is ruled out by cosmological observations. In the simplest case, only the NH with a low value of the lightest neutrino mass $m_1=0.01$ eV is considered in the following figures.
Fig. \ref{fig_nh_4neutrino_bj2} explains the minimum value of $\Delta m_{41}^2$ required by the measured value of $\langle m_{\beta\beta} \rangle$ from $0\nu$DBD experiments. For example, if the measurement of $\langle m_{\beta\beta} \rangle$ in an experiment is given by the vertical line, the allowed range of $m_4$ can be estimated from the intersection of the vertical and the shaded region in Fig. \ref{fig_nh_4neutrino_bj2}. The bound of $\Delta m_{41}^2$ naturally depends on $\sin^22\theta_{14}$, and so the following three values are examined: $\sin^22\theta_{14}= 0.01$, $0.10$ and $0.30$.

\begin{figure*}
\resizebox{170mm}{!}{\includegraphics[width=.80\textwidth]{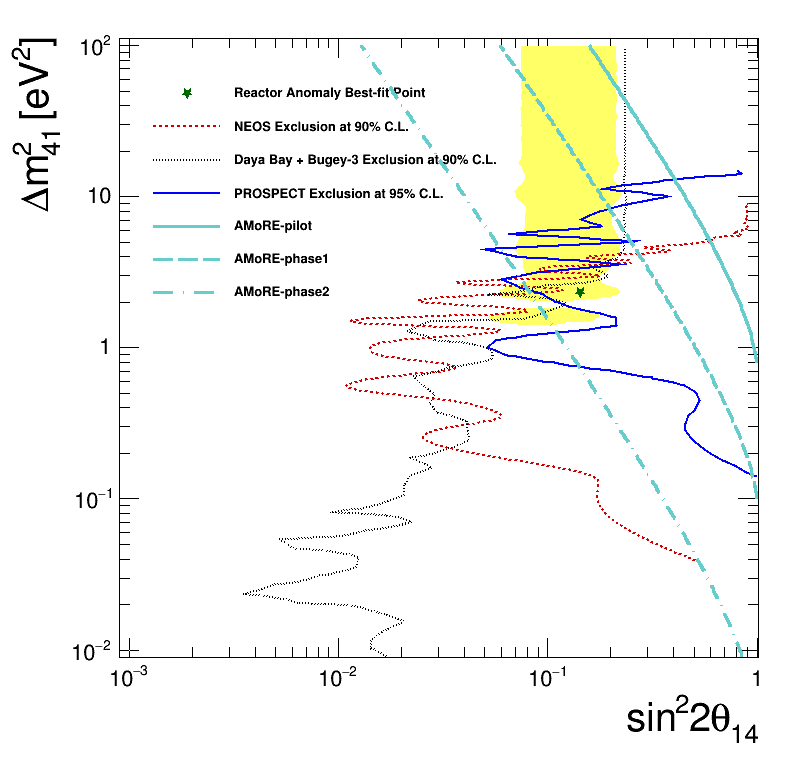}
                    \includegraphics[width=.80\textwidth]{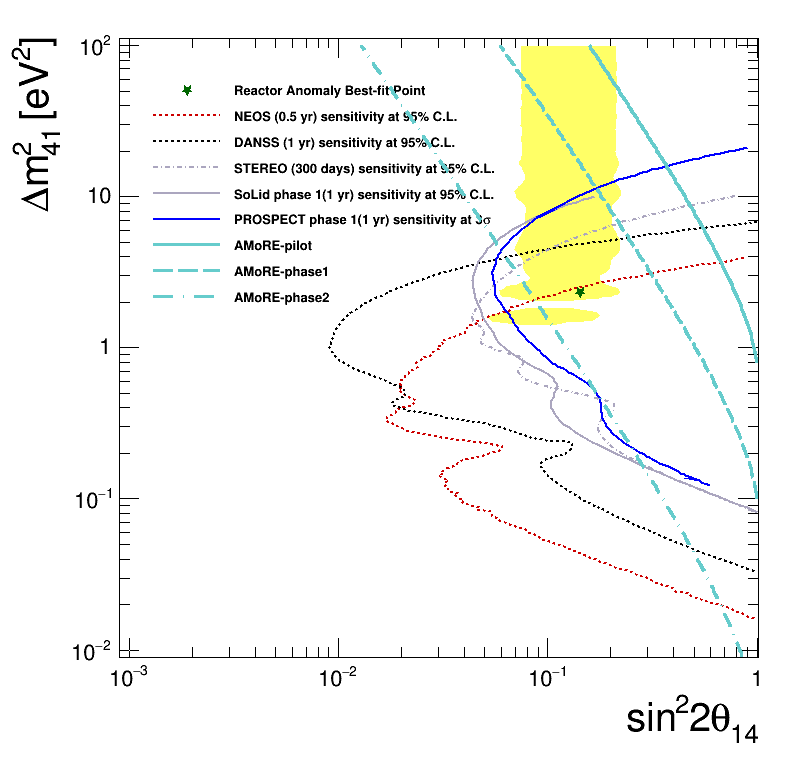}}
\caption{\label{fig_exclusion}
$\Delta m_{41}^2$ vs. $\sin^22\theta_{14}$ with the sensitivities of AMoRE experiments. The $\langle m_{\beta\beta} \rangle$ in Eq.(\ref{effective4m}) provides a limit in the $\Delta m_{41}^2$ and $\sin^22\theta_{14}$ plane. Neutrino masses are assumed to be of normal ordering. The curves cutting the right corners  indicate the sensitivities from AMoRE pilot(thick), phase 1(thin), and phase 2(dashed), respectively. The exclusion curves(left) of Daya Bay\cite{Adamson:2016jku}, NEOS\cite{Ko:2016owz}, and PROSPECT\cite{Ashenfelter:2018iov}, and the sensitivity curves(right) of SoLiD\cite{Abreu:2017bpe}, DANSS\cite{Alekseev:2018efk}, and STEREO\cite{Almazan:2018wln} are overlaid with AMoRE sensitivities. The yellow shades indicate $1\sigma$ region of the reactor antineutrino anomaly\cite{Mention:2011rk}.}
\end{figure*}

The correlation between $\langle m_{\beta\beta} \rangle$, $\Delta m_{41}^2$ and $\sin^22\theta_{14}$ guarantees that the measurement of $\langle m_{\beta\beta} \rangle$ can provide an exclusion contour in the $\Delta m_{41}^2$ and $\sin^22\theta_{14}$ plane. This way, improving the sensitivity to $0\nu$DBD can also be a stepwise strategy to scan the $\Delta m_{41}^2-\sin^22\theta_{14}$ space starting from the corner with heavier $m_4$ and larger mixing angles. Fig. \ref{fig_exclusion} shows that the sensitivities in neutrinoless double beta decay experiments can be considered together with exclusion contours or with sensitivities of short-baseline neutrino oscillation experiments. The best-fit and $1\sigma$ region for sterile neutrino expected by reactor anomaly are displayed for comparison\cite{Mention:2011rk}.

\section{Conclusion}

The results of the $0\nu$DBD search experiments are eagerly awaited to solve the problems of neutrino mass origin and the matter-antimatter asymmetry. However, if the neutrinos follow the normal mass ordering, one should find an experimental strategy which is much more sensitive than what is currently possible. In this work, we added a light sterile neutrino to the three neutrinos in the normal hierarchy, and predicted the increase in the effective electron-neutrino mass $\langle m_{\beta\beta} \rangle$. Accordingly, if the search is successful, it can  indirectly demonstrate the existence of light sterile neutrinos in addition to the validation of Majorana neutrinos. On the other hand, if several experiments in progress do not find any $0\nu$DBD events at a given sensitivity, some ranges of the parameters for the sterile neutrinos will be excluded, if neutrinos are Majorana particles. It has been shown that narrowing the sterile-neutrino mass and mixing angle range by measuring the effective mass of the virtual neutrino between the double-beta decay vertices can compensate the limitations from short-baseline oscillation experiments.

\begin{acknowledgments}
This work was supported by the National Research Foundation Grant of Korea (NRF-2017R1A2B4004308).
\end{acknowledgments}


\begin{thebibliography} {99}

\bibitem{Giunti_Kim:2009}
	C. Giunti, C. W. Kim. 2009. Fundamentals of neutrino physics and astrophysics, Oxford Press.
\bibitem{Vergados:2002pv}
  J.~D.~Vergados,
  Phys.\ Rept.\  {\bf 361}, 1 (2002)
\bibitem{Barea:2012zz}
  J.~Barea, J.~Kotila and F.~Iachello,
  Phys.\ Rev.\ Lett.\  {\bf 109}, 042501 (2012)
\bibitem{Bilenky:2012qi}
  S.~M.~Bilenky and C.~Giunti,
  Mod.\ Phys.\ Lett.\ A {\bf 27}, 1230015 (2012)
\bibitem{lep} M. Fukugita and T. Yanagida, Phys. Lett. {\bf B 174}, 45 (1986).
\bibitem{Harvey:1990qw}
J.~A.~Harvey and M.~S.~Turner, Phys.\ Rev.\ D {\bf 42}, 3344
(1990).
  H.~B.~Nielsen and Y.~Takanishi,
  Phys.\ Lett.\ B {\bf 507}, 241 (2001)
\bibitem{Kolb:qa}
E.~W.~Kolb and S.~Wolfram,
Nucl.\ Phys.\ B {\bf 172}, 224 (1980) [Erratum-ibid.\ B {\bf 195},
542 (1982)].
\bibitem{Barger:2003gt}
  V.~Barger, D.~A.~Dicus, H.~J.~He and T.~j.~Li,
  Phys.\ Lett.\ B {\bf 583}, 173 (2004)
\bibitem{Luty:un}
M.~A.~Luty,
Phys.\ Rev.\ D {\bf 45}, 455 (1992).
Phys.\ Lett.\ B {\bf 345}, 248 (1995) [Erratum-ibid.\ B {\bf 382},
447 (1996)]
L.~Covi, E.~Roulet and F.~Vissani,
Phys.\ Lett.\ B {\bf 384}, 169 (1996)
W.~Buchmuller and M.~Plumacher,
Phys.\ Lett.\ B {\bf 431}, 354 (1998)
W.~Buchmuller and M.~Plumacher,
Int.\ J.\ Mod.\ Phys.\ A {\bf 15}, 5047 (2000)
\bibitem{Endoh:2002wm}
  T.~Endoh, S.~Kaneko, S.~K.~Kang, T.~Morozumi and M.~Tanimoto,
  Phys.\ Rev.\ Lett.\  {\bf 89}, 231601 (2002)
  S.~Davidson and A.~Ibarra,
  Nucl.\ Phys.\ B {\bf 648}, 345 (2003)
  G.~C.~Branco, R.~Gonzalez Felipe, F.~R.~Joaquim, I.~Masina, M.~N.~Rebelo and C.~A.~Savoy,
  Phys.\ Rev.\ D {\bf 67}, 073025 (2003)
  A.~de Gouvea, B.~Kayser and R.~N.~Mohapatra,
  Phys.\ Rev.\ D {\bf 67}, 053004 (2003)
  S.~Pascoli, S.~T.~Petcov and W.~Rodejohann,
  Phys.\ Rev.\ D {\bf 68}, 093007 (2003)
  W.~Grimus and L.~Lavoura,
  J.\ Phys.\ G {\bf 30}, 1073 (2004)
  A.~Ibarra and G.~G.~Ross,
  Phys.\ Lett.\ B {\bf 591}, 285 (2004)
  S.~Davidson and R.~Kitano,
  JHEP {\bf 0403}, 020 (2004)
  M.~C.~Chen and K.~T.~Mahanthappa,
  Phys.\ Rev.\ D {\bf 71}, 035001 (2005)
  S.~Pascoli, S.~T.~Petcov and A.~Riotto,
  Nucl.\ Phys.\ B {\bf 774}, 1 (2007)
\bibitem{Siyeon:2016wro}
  K.~Siyeon,
  J.\ Korean Phys.\ Soc.\  {\bf 69}, no. 11, 1638 (2016)
\bibitem{KamLAND-Zen:2016pfg}
  A.~Gando {\it et al.} [KamLAND-Zen Collaboration],
  Phys.\ Rev.\ Lett.\  {\bf 117}, no. 8, 082503 (2016)
  Addendum: [Phys.\ Rev.\ Lett.\  {\bf 117}, no. 10, 109903 (2016)]
\bibitem{Albert:2017owj}
  J.~B.~Albert {\it et al.} [EXO Collaboration],
  Phys.\ Rev.\ Lett.\  {\bf 120}, no. 7, 072701 (2018)
\bibitem{Alduino:2017ehq}
  C.~Alduino {\it et al.} [CUORE Collaboration],
  Phys.\ Rev.\ Lett.\  {\bf 120}, no. 13, 132501 (2018)
\bibitem{Agostini:2018tnm}
  M.~Agostini {\it et al.} [GERDA Collaboration],
  Phys.\ Rev.\ Lett.\  {\bf 120}, no. 13, 132503 (2018)
\bibitem{Alvis:2018yte}
  S.~I.~Alvis {\it et al.} [Majorana Collaboration],
  Phys.\ Rev.\ Lett.\  {\bf 120}, no. 21, 211804 (2018)
\bibitem{Arnold:2018tmo}
  R.~Arnold {\it et al.},
  Eur.\ Phys.\ J.\ C {\bf 78}, no. 10, 821 (2018)
\bibitem{Alenkov:2015dic}
  V.~Alenkov {\it et al.} [AMoRE Collaboration],
 arXiv:1512.05957 [physics.ins-det].
\bibitem{Lee:2016pmi}
  J.~Y.~Lee {\it et al.},
  IEEE Trans.\ Nucl.\ Sci.\  {\bf 63}, no. 2, 543 (2016).
\bibitem{Luqman:2016okt}
  A.~Luqman {\it et al.},
  Nucl.\ Instrum.\ Meth.\ A {\bf 855}, 140 (2017)
\bibitem{Adamson:2017gxd}
  P.~Adamson {\it et al.} [NOvA Collaboration],
  Phys.\ Rev.\ Lett.\  {\bf 118}, no. 23, 231801 (2017)
\bibitem{Abe:2018wpn}
  K.~Abe {\it et al.} [T2K Collaboration],
  Phys.\ Rev.\ Lett.\  {\bf 121}, 171802 (2018)
\bibitem{Adamson:2016jku}
  P.~Adamson {\it et al.} [Daya Bay and MINOS Collaborations],
  Phys.\ Rev.\ Lett.\  {\bf 117}, no. 15, 151801 (2016)
  Addendum: [Phys.\ Rev.\ Lett.\  {\bf 117}, no. 20, 209901 (2016)]
\bibitem{Ko:2016owz}
  Y.~J.~Ko {\it et al.} [NEOS Collaboration],
  Phys.\ Rev.\ Lett.\  {\bf 118}, no. 12, 121802 (2017)
\bibitem{Ashenfelter:2018iov}
  J.~Ashenfelter {\it et al.} [PROSPECT Collaboration],
  arXiv:1806.02784 [hep-ex].
\bibitem{Abreu:2017bpe}
  Y.~Abreu {\it et al.} [SoLid Collaboration],
  JINST {\bf 12}, no. 04, P04024 (2017)
  \bibitem{Alekseev:2018efk}
  I.~Alekseev {\it et al.} [DANSS Collaboration],
  Phys.\ Lett. \ B (2018), no. 10, 038
\bibitem{Almazan:2018wln}
  H.~Almazán {\it et al.} [STEREO Collaboration],
  Phys.\ Rev.\ Lett.\  {\bf 121}, no. 16, 161801 (2018)
\bibitem{Elliott:1987kp}
  S.~R.~Elliott, A.~A.~Hahn and M.~K.~Moe,
  Phys.\ Rev.\ Lett.\  {\bf 59}, 2020 (1987).
\bibitem{Barabash:2010ie}
  A.~S.~Barabash,
  Phys.\ Rev.\ C {\bf 81}, 035501 (2010)
\bibitem{Saakyan:2013yna}
  R.~Saakyan,
  Ann.\ Rev.\ Nucl.\ Part.\ Sci.\  {\bf 63}, 503 (2013).
\bibitem{Faessler:2012ku}
  A.~Faessler, V.~Rodin and F.~Simkovic,
  J.\ Phys.\ G {\bf 39}, 124006 (2012)
\bibitem{Tanabashi:2018oca}
  M.~Tanabashi {\it et al.} [Particle Data Group],
  Phys.\ Rev.\ D {\bf 98}, no. 3, 030001 (2018).
\bibitem{Drexlin:2005zt}
  G.~Drexlin [KATRIN Collaboration],
  Nucl.\ Phys.\ Proc.\ Suppl.\  {\bf 145}, 263 (2005).
\bibitem{Giunti:2015kza}
  C.~Giunti and E.~M.~Zavanin,
  JHEP {\bf 1507}, 171 (2015)
\bibitem{Mention:2011rk}
  G.~Mention, M.~Fechner, T.~Lasserre, T.~A.~Mueller, D.~Lhuillier, M.~Cribier and A.~Letourneau,
  Phys.\ Rev.\ D {\bf 83}, 073006 (2011)
  \end{thebibliography}
\end{document}